\documentclass{article}

\usepackage{arxiv}

\usepackage[utf8]{inputenc} % allow utf-8 input
\usepackage[T1]{fontenc}    % use 8-bit T1 fonts
\usepackage[hidelinks]{hyperref}
\usepackage{url}            % simple URL typesetting
\usepackage{booktabs}       % professional-quality tables
\usepackage{tabularx}
\usepackage{amsfonts}       % blackboard math symbols
\usepackage{nicefrac}       % compact symbols for 1/2, etc.
\usepackage{microtype}      % microtypography
\usepackage{lipsum}
\usepackage{graphicx}
\graphicspath{ {./images/} }

\title{TruEyes: Utilizing Microtasks in Mobile Apps for Crowdsourced Labeling of Machine Learning Datasets}

\author{
 Chandramohan Sudar \\
  Center for Digital Technology and Management, Germany \\
  Technical University of Munich, Germany \\
  \texttt{chandramohan.sudar@cdtm.de} \\
  \And
 Michael Froehlich \\
  Center for Digital Technology and Management, Germany \\
  Ludwig Maximilian University, Germany \\
  Bundeswehr University Munich, Germany \\
  \texttt{froehlich@cdtm.de} \\
  \And
 Florian Alt \\
  Bundeswehr University Munich, Germany \\
  \texttt{florian.alt@unibw.de} \\
}

\begin{document}
\maketitle
\begin{abstract}
The growing use of supervised machine learning in research and industry has increased the need for labeled datasets.
Crowdsourcing has emerged as a popular method to create data labels.
However, working on large batches of tasks leads to worker fatigue, negatively impacting labeling quality.
To address this, we present TruEyes, a collaborative crowdsourcing system, enabling the distribution of micro-tasks to mobile app users.
TruEyes allows machine learning practitioners to publish labeling tasks, mobile app developers to integrate task ads for monetization, and users to label data instead of watching advertisements. 
To evaluate the system, we conducted an experiment with N=296 participants.
Our results show that the quality of the labeled data is comparable to traditional crowdsourcing approaches and most users prefer task ads over traditional ads. 
We discuss extensions to the system and address how mobile advertisement space can be used as a productive resource in the future.

% keywords can be removed
\end{abstract}

\section{Introduction}

Over the last decade, there has been a rapid increase in the adoption of Machine Learning (ML) in various application areas \cite{jo15}.
As research and industry use cases are constantly evolving, there is a growing demand for high-quality labeled data to train and evaluate new ML models.
While automated label generation techniques have been advancing, in many cases it still requires humans to reliably annotate data \cite{ash21}.
Depending on the nature of the labeling task and the size of the dataset, data labeling can quickly become expensive and time-consuming \cite{kas21}.
Crowdsourcing platforms have become a popular method to outsource labeling tasks at relatively low costs \cite{ash21}.
Platforms such as Amazon Mechanical Turks\footnote{\href{https://www.mturk.com/}{https://www.mturk.com/} (last-accessed 2022-04-21)} (MTurk) have been widely used in various stages of ML including data labeling \cite{va17}.
However, these platforms are not without challenges in terms of quality, cost and latency \cite{alo15} and bring along the risk of treating workers as computing resources rather than recognizing the humans behind the platform \cite{Barbosa2019Rehumanizing}.
Particularly labeling large sets of data comes with challenges:
The monotonous and repetitive nature of tasks can result in workers losing motivation resulting in a reduced quality of the labeled data \cite{mai21,kas21}.
It is not surprising to imagine why – research shows that working on large batches of tasks leads to both physical and cognitive fatigue, impacting not only workers' well-being but also leading to less reliable output in the process \cite{Dai2015MicroDiversions, Krueger1989SustainedWork}.

We address this problem by exploring a novel approach to mobile crowdsourcing.
Instead of completing labeling tasks in large batches, we distribute them to mobile app users in the form of micro-tasks that take no longer than 30 seconds to complete.
Several studies indicate the 20-30 second mark to be ideal for completing short surveys and questionnaires \cite{ma14}. 
To this end, we developed \textit{TruEyes}, a crowdsourcing system that leverages mobile app users to perform data labeling tasks that are typically distributed to crowdsourcing platforms.
The end-user experience and the content delivery approach is similar to mobile ads, in particular interstitials ad formats\footnote{For more information see: \url{https://support.google.com/admob/answer/6066980} (last-accessed: 2022-04-21)}.
Interstitial ads are designed to be placed between content as they partially or completely disrupt the interaction with the host app while being displayed.
These ads are typically served at natural transition points in the flow of an app, e.g., during the pause between levels in a game \cite{ngu20}.
When the host app is ready to show an interstitial ad, the system injects a set of data labeling tasks and records the user response.

The proposed system is referred to as TruEyes in the rest of this document. The data labeling task visible to a mobile app user is referred to as a task ad. 
TruEyes Console is the subsystem that manages the labeling dataset, creates the labeling tasks, and consolidates the validated labels.
TruEyes Client is the subsystem that distributes labeling tasks to mobile app users and records the user response.
TruEyes Software Development Kit (SDK) is the subsystem integrating with the host app to orchestrate 
when the labeling tasks is shown.

To evaluate the developed system and test the efficiency of our approach, we integrated the solution into a mobile game and and conducted a randomized online experiment with N=296 participants.
We tested two incentive settings: One in which the participants had the choice to perform a task ad in return for game points as reward, and another in which the task ads were non-optional and could not be skipped before returning to the game. 
Our implementation was successful in delivering the task ads and capturing labeling responses for the dataset.
The accuracy of the resulting datasets is comparable to a control group using mTurk: We observed a median success rate of 80\% for the rewarded setting and 84\% for the non-optional setting, compared to a median success rate of 82\% in our control group.
A post-experiment survey of participants further indicated a positive sentiment for task ads when compared to regular mobile ads. 
We discuss implications arising from these results for using microtasks for data labeling, particularly in how far they can become a viable alternative to existing approaches with regards to cost, quality, and participant diversity.

\vspace{1mm}
\noindent
\textbf{Contribution Statement: }
In this paper we present two primary contributions.
First, we outline an architectural approach for realizing micro-task based data labeling utilizing mobile app users. 
Second, we provide an empirical evaluation of the implemented system with N=296 participants assessing the feasibility of the proposed approach. 
In line with the theme of NordiCHI 2022 we believe that micro-tasks distributed through mobile apps can be an effective participatory way to create high-quality labels for machine learning datasets while also recognizing the human needs and well-being of crowdsourced workers.

\section{Background \& Related Work}

% What is Crowdsourcing?
Crowdsourcing is the act of outsourcing a task to a large network of people \cite{yu12}. The ubiquity of the Internet has played a major role in the success of crowdsourcing, by enabling tasks to be distributed on a globe scale \cite{br08}. Due to its popularity and wide adoption in solving various industry use cases, it is now considered an indispensable computational resource \cite{da15}. Paid-crowdsourcing platforms, such as MTurk, provide workers the opportunity to participate in tasks for monetary compensation. In recent times, MTurk has experienced wide adoption in research communities to collect data and validate results \cite{ga15}. Such crowdsourcing projects have contributed novel datasets that have been used for training machine learning models \cite{de09, kr17, po12}.

Although workers in these platforms participate for monetary compensation, higher rewards do not necessarily mean better quality results \cite{fey15}. 
In addition to extrinsic motivation such as monetary compensation, maximizing intrinsic motivation can lead to improved task performance \cite{mas09}. 
Many workers prefer tasks that are engaging and intrinsically rewarding over tasks that offer purely monetary rewards \cite{eic12}. However, it can be extremely challenging to redesign existing tasks such that they are more engaging to crowdsourced workers. 

Another challenge associated with crowdsourcing is its monotonous nature experienced by workers when performing large batches of tasks. In paid-crowdsourcing platforms such as MTurk, workers may switch between tasks in search of more engaging work. Hence, requesters risk losing experienced workers who produce high quality results. Dai et al. \cite{da15} studied the effects of incorporating micro-diversions such as games and comics in between long stretches of monotonous crowdsourcing tasks in order to retain workers. Their findings indicated an increase in worker retention and speed, while maintaining the same quality of work.

\subsection{Mobile Crowdsourcing}
Mobile crowdsourcing makes use of mobile devices to engage participants to perform crowdsourcing tasks. The increasing use of smart mobile devices along with the added convenience of performing tasks irrespective of time and location in a handheld device has largely contributed to the rise of mobile crowdsourcing \cite{ik17, gu12, ea09}. Over the years, various mobile crowdsourcing platforms and services have emerged, some of them taking advantage of the sensors present in the mobile device including GPS \cite{na19}. Notably, Waze\footnote{See \url{https://waze.com/} (last-accessed: 2022-04-21)} is an example of such a service that relies on crowdsourced data for traffic monitoring and incident reporting to enhance its real-time navigation service \cite{am18}. 

In contrast to traditional crowdsourcing platforms, mobile crowdsourcing has been able to achieve worker participation for non-monetary incentives. The Crowdsource app\footnote{See \url{https://crowdsource.google.com/about/} (last-accessed: 2022-04-21)}, developed by Google, leverages its users to acquire and improve training data for machine learning projects \cite{ch18}. Users can complete various types of microtasks such as image classification, audio validation, translation validation etc. Instead of monetary compensation, it gamifies tasks by incorporating rewards in the form of points, levels and badges \cite{ch18}. Another key distinction employed by mobile crowdsourcing is the increased use of microtasks. Studies have shown that traditional crowdsourcing campaigns often fail due to lack of time and motivation for the participants \cite{po09, da11}. Twitch Crowdsourcing explored microtasks in mobile crowdsourcing by requesting users to complete a microtask everytime they unlocked their mobile device \cite{va14}. This provides the opportunity to engage users in their spare time for "ad-hoc" crowdsourcing.  With this approach, participants were able to complete microtasks such as collecting contextual data, rating stock photos and help structure information from the web \cite{va14}.

The ad-hoc approach to crowdsourcing was further extended by CrowdPickUp \cite{go17}, which combines mobile crowdsourcing with situated crowdsourcing, a technique that makes use of situated technologies such as public displays, kiosks and other interactive input mechanisms to recruit users and perform crowdsourcing tasks \cite{ho14}. CrowdPickUp provides a web platform that allows participants to engage in crowdsourcing tasks using their mobile device. To access the platform, participants either scan a QR code or visit a shortened URL that is displayed at physical locations \cite{go17}. The use of a web platform that does not require users to download and install an additional application to access the tasks lowers the barrier of entry for participation. The participants are further incentivized by rewards in the form of virtual coins that can be redeemed in exchange for real money or movie tickets.

It is evident from the findings of Twitch Crowdsourcing \cite{va14}, and CrowdPickUp \cite{go17} that, day-to-day  users can be leveraged to perform crowdsourcing tasks in their mobile devices. However, they still rely on workers actively volunteering or signing up to perform tasks in one way or another. Thus, recruiting ad-hoc workers can become the primary obstacle when deploying such solutions at scale. Moreover, as with traditional crowdsourcing, the recruited workers still need to be retained. Techniques such as micro-diversion proposed by Dai et al. can be adapted for the mobile crowdsourcing workflow to achieve this.

\subsection{Summary}
Building on the existing body of literature we can derive several implications for exploring the use of mobile ad space for microtasks.
Extending existing mobile crowdsourcing approaches from literature, we consider the following conditions and requirements for the design and implementation of the presented system:

\begin{itemize}
    \item \textbf{Microtasks}: Individual workers should make small contributions instead of batch tasks leading to worker fatigue.
    \item \textbf{Worker Engagement}: Tasks should be delivered to worker in an engaging way, i.e., as interactive diversions.
    \item \textbf{Ad-hoc Availability}: Workers can be recruited on-demand for performing tasks.
    \item \textbf{Incentive System}: Workers need to be incentivized to perform tasks.
\end{itemize}

\noindent
We believe that mobile advertisement (ad) as a medium can satisfy all the above conditions to provide a suitable crowdsourcing environment. Mobile ads are the major source of revenue for mobile apps\footnote{See \url{https://statista.com/topics/983/mobile-app-monetization/} (last-accessed: 2022-04-21)}. 
The smaller form factor of mobile devices result in ads partially or fully occupying the user interface of the host app, referred to as banner ads and interstitial ads respectively.
Interstitial ads can be considered as diversions during the usage of the host app since they can last for up to 30 seconds and are generally shown only at natural breakpoints in the host app.
This duration is ideal for performing a microtask.
As the primary activity of the user is engaging with the host app, we argue that our approach reaps the benefits of micro-diversion proposed by Dai and Rzeszotarski \cite{da15}, with the difference being that the crowdsourcing task now becomes the micro-diversion during the host app usage.
In recent times, ad publishers have introduced new interstitial ad formats such as rewarded ads and interactive playable ads to further incentivize users to engage with ads.
Rewarded ads offer users in-app rewards in return for engaging with an ad, while interactive playable ads offer interactive content such as mini-games playable within the ad, which further validates the technical viability of deploying interactive microtasks and also provides opportunities to incorporate reward based incentives for completing microtasks.
Moreover, the on-demand nature of mobile ads and the access to a large pool of mobile app users further satisfies our proposed conditions for creating an improved mobile crowdsourcing sytem. 
However, it is necessary to understand if mobile app users have enough incentive to perform these tasks and analyze whether their performance levels are satisfactory for real-world data labeling needs.

\section{TruEyes}

In order to conceptualize our proposed mobile crowdsourcing approach, we first identified the primary users of the system and their respective roles.
By defining a real-life crowdsourcing scenario with the identified users, we gathered functional and non-functional requirements based on the necessary interactions with the system \cite{bru09}.
This allowed us to clearly separate concerns of the underlying architecture and define the user interfaces. 

\subsection{Actors and Interaction Flow}
We identified three primary users that interact directly with the TruEyes system, namely, ML practitioners, app users and app developers.
The \textsc{ML Practitioner} is interested in crowdsourcing labels for their ML project. They want to store their labeled and unlabeled datasets. They should be able to design and preview the labeling task. They need to publish the labeling task and set their quality preference. ML practitioners require tooling to manage labeling datasets, design tasks with interactive builders and publishing parameters to adjust the labeling quality.
The \textsc{App User} is responsible for labeling the data. They interact directly with the labeling user interface. Their primary activity is engaging with the host app. When a task ad is shown over the host app, they should be able to complete the labeling task and return to the host app. App users require instructions for the labeling task, an interface to perform the labeling task and return to the host app.
The \textsc{App Developer} wants to create an engaging secondary experience for the app user in order to monetize their app. Similar to traditional mobile ads, app developers must be able to invoke the TruEyes task ads to embed the labeling tasks on demand. App developers require a programming interface with the system that allows them to add labeling task ads in between their app sessions.

\begin{figure*}[h]
  \vspace{-2mm}
  \centering
  \includegraphics[width=0.9\linewidth]{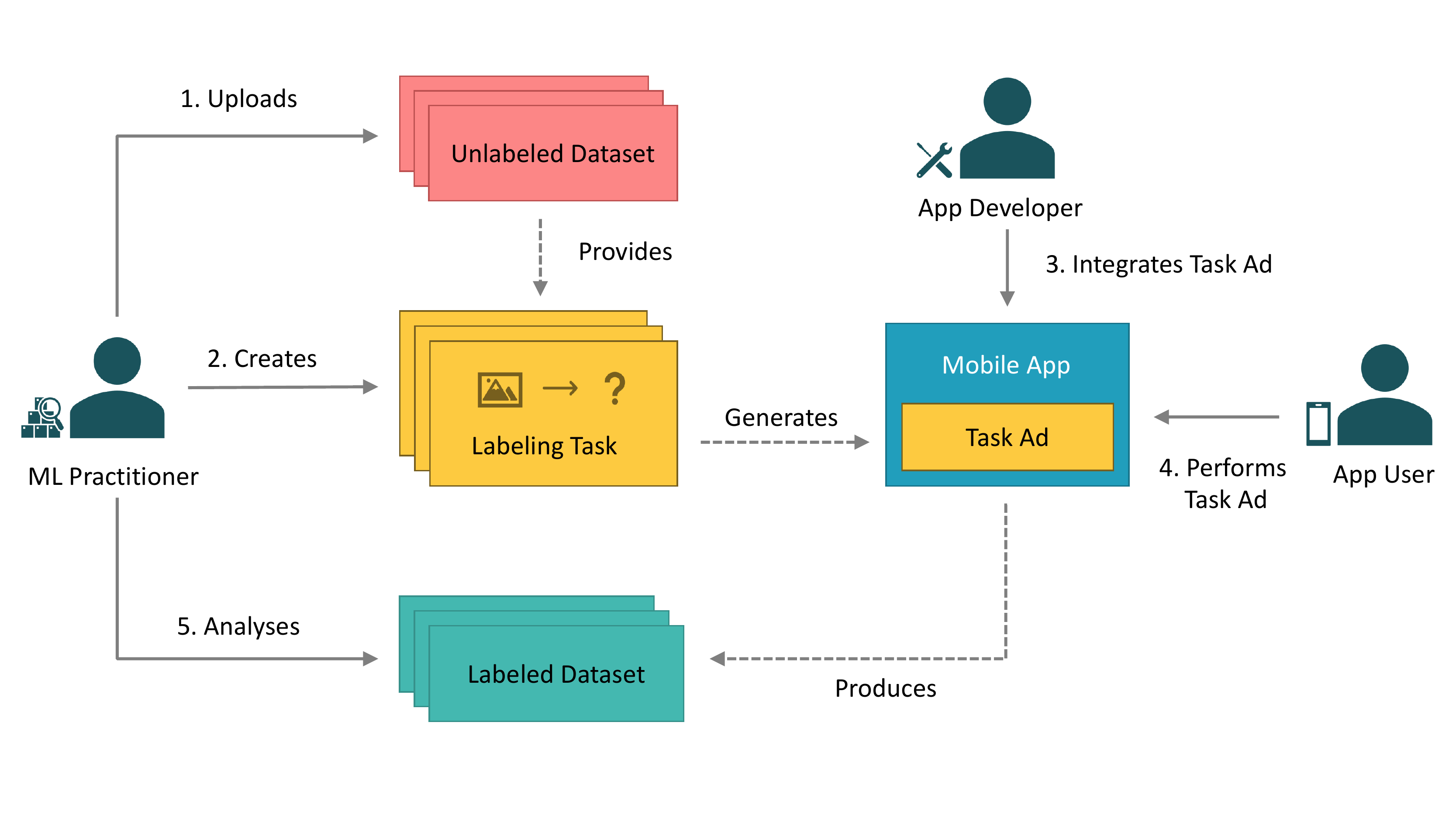}
  \vspace{-2mm}
  \caption{Interaction flow between the users of the TruEyes system.}
  \label{fig:interactionflow}
  \vspace{-2mm}
\end{figure*}

\noindent
Figure \ref{fig:interactionflow} illustrates the use cases through which each actor interacts with the proposed system and provides an overview of the interaction flow:
First, the \textsc{ML practitioner} uploads the unlabeled dataset and creates the labeling tasks. 
Independent \textsc{app developers} can then implement the framework in their apps and hand over the control flow to the system at specific points in their app to invoke the labeling task ads. 
When the \textsc{app users} are presented the task ad, they perform the labeling task.
Once finished, the control flow is handed back to the host app.
Through the completion of task ads by different users a labeled dataset is constructed over time.
The \textsc{ML practitioner} can view and analyze the labeled dataset collected through the labeling tasks.

\subsection{Conceptual Architecture}
\begin{figure}[h!]
  \centering
  \vspace{-2mm}
  \includegraphics[width=0.9\linewidth]{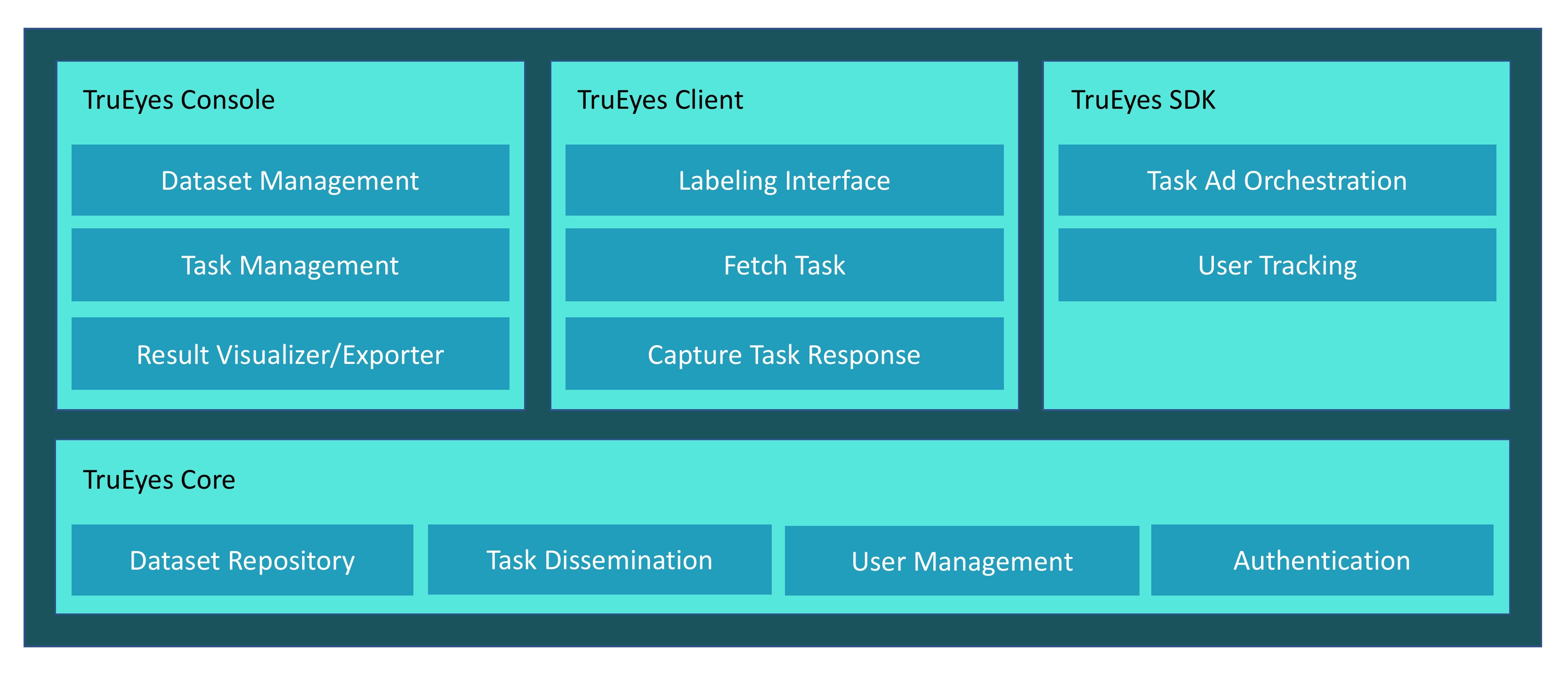}
  \vspace{-1mm}
  \caption{High-level design of the TruEyes system.}
  \label{fig:highleveldesign}
  \vspace{-2mm}
\end{figure}

The proposed system is broken down into three subsystems namely, TruEyes Console, TruEyes Client, and TruEyes SDK.
Figure \ref{fig:highleveldesign} depicts the high-level design of the system and its subsystems.

\vspace{1mm}
\noindent
\textbf{TruEyes Console} provides a content management system (CMS) for managing the labeling workflow. It allows ML practitioners to upload their unlabeled dataset, which is then pre-processed and stored in a format that is accessible by the internal subsystems. It provides an interface for ML practitioners to input labeling instructions and customize the labeling interface. Parameters such as the number of labeling workers can be tweaked in order to adjust the quality of labels. Finally, it provides the option to publish and unpublish the labeling tasks, with the ability to view and analyze the results of the labeling tasks. 

\vspace{1mm}
\noindent
\textbf{TruEyes SDK} provides an interface for mobile app developers to invoke the TruEyes task ads at specific instances in the user flow of the app. After a task ad is invoked, the program flow of the app is transferred to the TruEyes SDK. The SDK requests the TruEyes Client to provide a set of  labeling tasks. When the user completes the labeling tasks or if the labeling tasks are not available, the program flow is transferred back to the app. Additionally, the SDK preserves user information to avoid serving labeling tasks that have already been completed by the user. 

\vspace{1mm}
\noindent
\textbf{TruEyes Client} is responsible for serving labeling tasks. First, a request for serving labeling tasks is received. The client then looks for the published labeling tasks that have not received the required number of worker engagements. After receiving the tasks, it uses the instructions and customizations set by the ML practitioner to render a user interface for mobile app users. It captures the response generated by the mobile app user along with usage metrics for each label.
 
\vspace{1mm}
\noindent
\textbf{TruEyes Core} brings together all the subsystems by providing a communication interface to trigger various actions and a data repository to store all the data. It is responsible for task dissemination, which involves creating labeling tasks using the dataset provided by the ML practitioner and distributing the tasks to TruEyes Clients based on the defined quality thresholds, ensuring that an App User does not view the same labeling task twice. Additionally, it handles utility services such as authentication and user management.

\subsection{Implementation}
In terms of the technical implementation, TruEyes Console was implemented using the React\footnote{\url{https://reactjs.org/} (last-accessed 2022-04-21)} javascript front-end library, deployed as a client-side rendering web app. 
The TruEyes Client made use of Preact\footnote{\url{https://preactjs.com/} (last-accessed 2022-04-21)}, a minimal alternative to React, deployed as a server-side rendering web app. 
The TruEyes SDK was developed as a native SDK for the Android\footnote{\url{https://www.android.com/}  (last-accessed 2022-04-21)} mobile platform, it renders the TruEyes Client as a native web app on Android using a WebView\footnote{\url{https://developer.android.com/reference/android/webkit/WebView}  (last-accessed 2022-04-21)}.
Decoupling the task rendering UI from the native SDK allowed us to push updates to the TruEyes Client without having to update the SDK. 
In practice, this ensures a seamless integration for App Developers as they do not have to manually update the SDK and create new releases for their mobile apps. 
All the subsystems were brought together by TruEyes Core, the backend system that handles all the communication via REST APIs using the Django\footnote{\url{https://www.djangoproject.com/} (last-accessed 2022-04-21)} web framework. 
We used a monolithic architecture to handle the backend services, which in contrast to a micro-service architecture is easier to develop and deploy \cite{go20}. 
All the subsystems were deployed on Amazon Web Services\footnote{\url{https://aws.amazon.com/}  (last-accessed 2022-04-21)}.

 \begin{figure*}[!h]
   \vspace{0mm}
   \centering
    \includegraphics[width=0.49\textwidth]{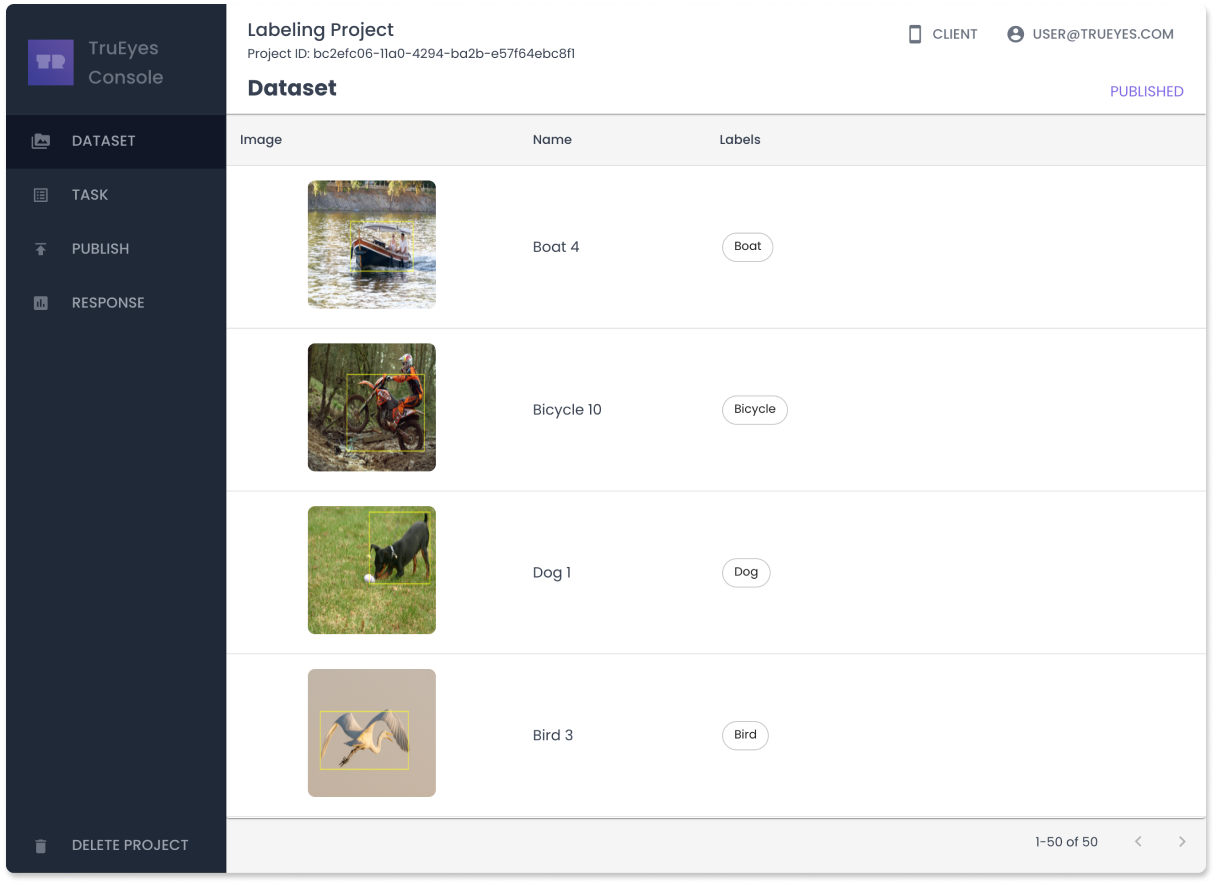}
    \includegraphics[width=0.49\textwidth]{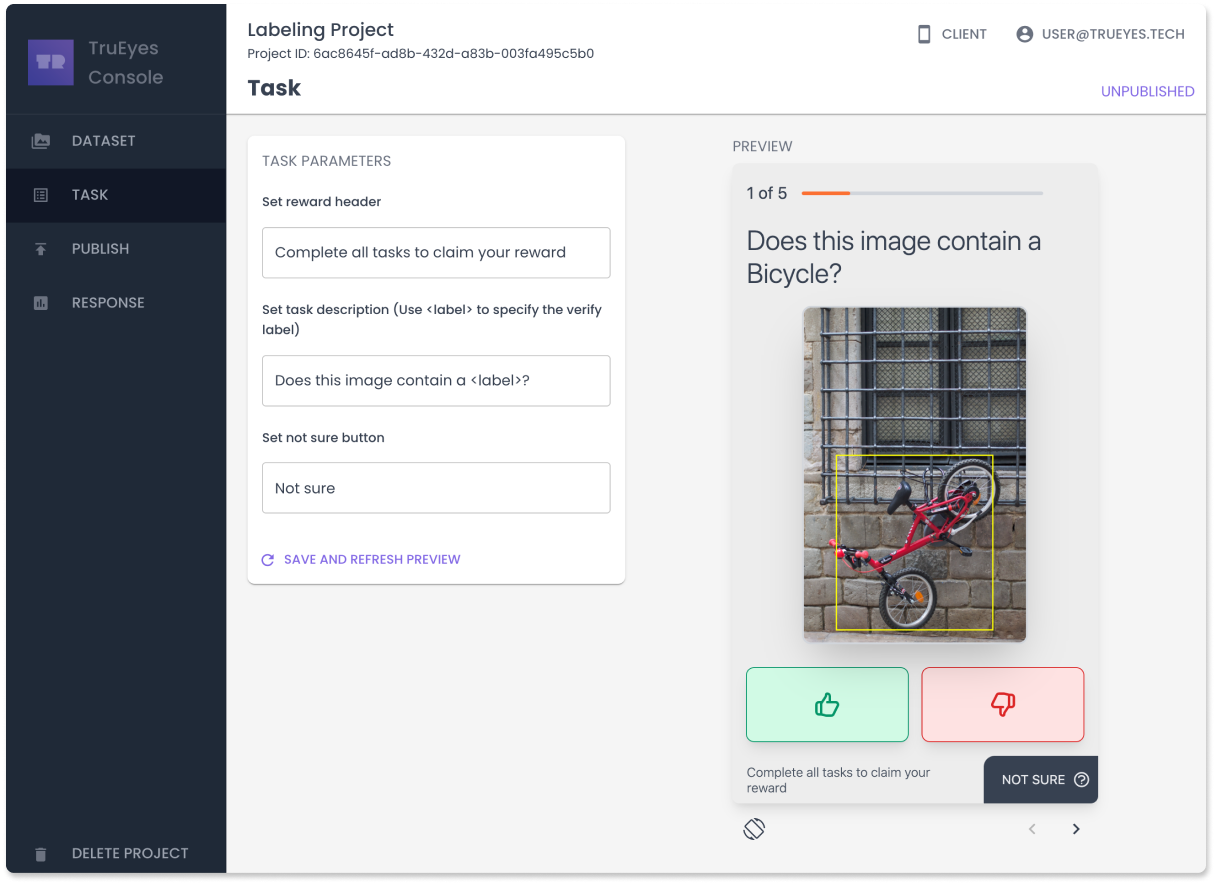}
    \includegraphics[width=0.49\textwidth]{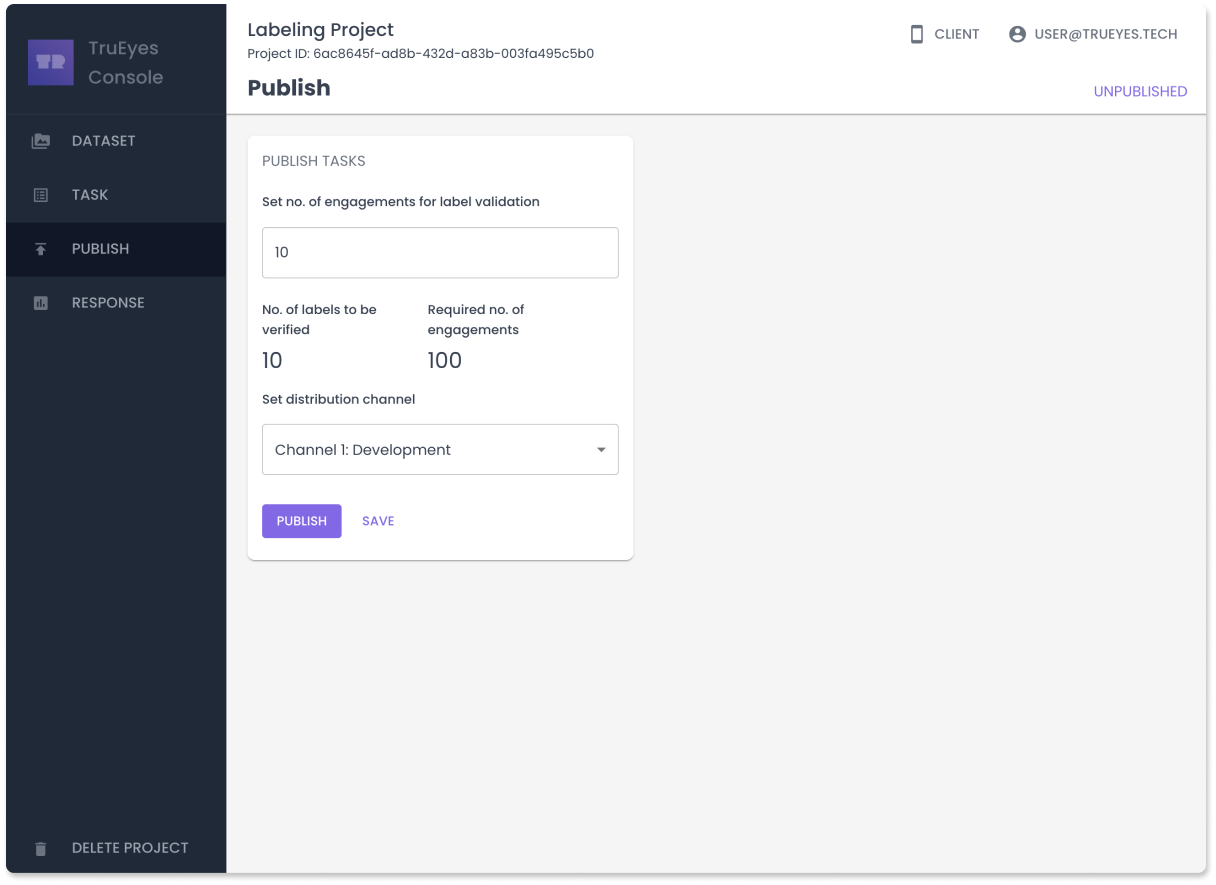}
    \includegraphics[width=0.49\textwidth]{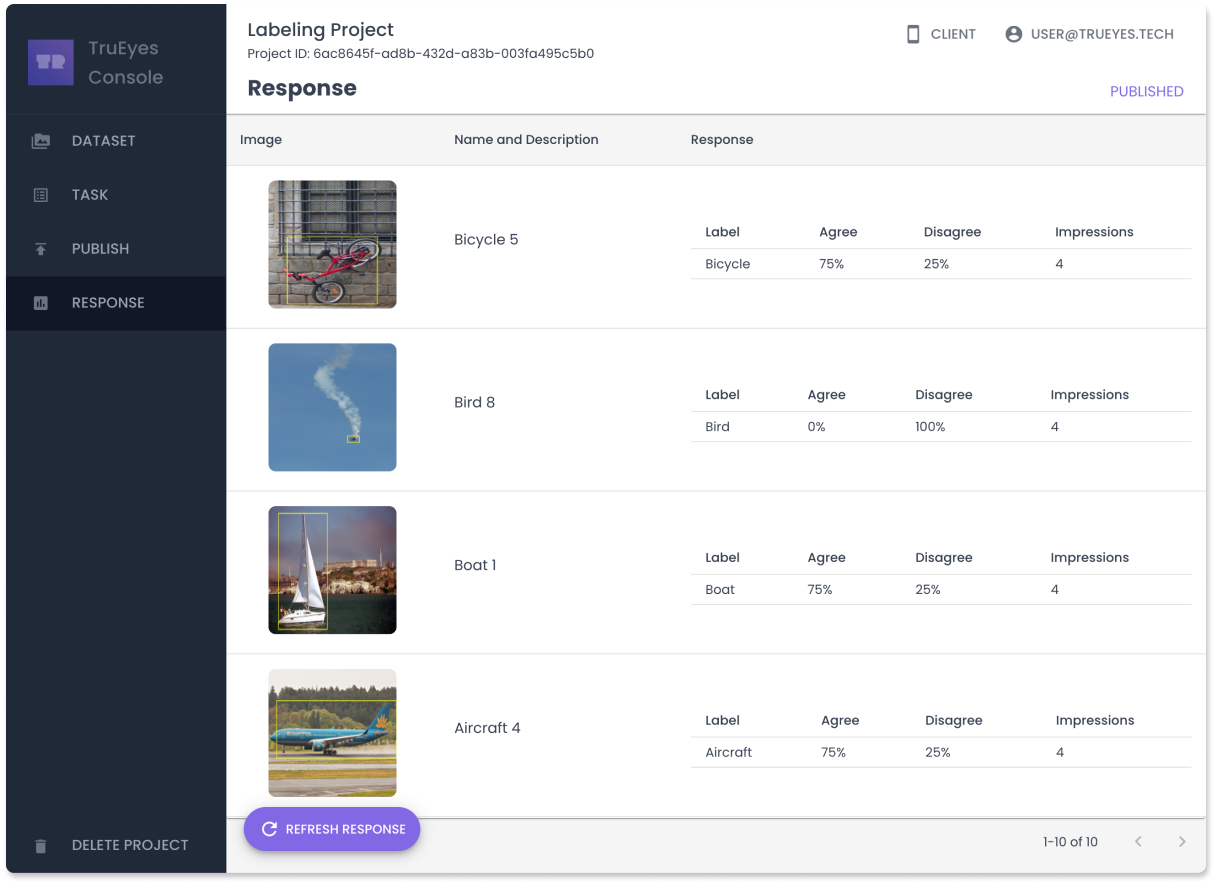}
   \vspace{-0mm}
   \caption{
    Interfaces of the implemented prototype for configuring labeling tasks. ML practitioners can upload new datasets (top left), customize and preview interfaces for the labeling task (top right), configure label quality parameters (bottom left), and review the the generated labels once the task is published (bottom right).
   }
   \label{fig:interfaces}
   \vspace{0mm}
\end{figure*}

\section{Evaluation}

We evaluated the implemented system with an online experiment. 
The goal of the evaluation was twofold: 
First, to demonstrate the feasibility of the overall approach and compare the data quality to a baseline scenario.
Second, to evaluate the system performance and collect data on how users would perceive task ads.

\subsection{Study Design}
We used a randomized online experiment with three conditions.
The independent variables are the type of labeling task delivery [mturk, rewarded task-ad, non-optional task ad]. 
The dependent variables are the success rate (percentage of correctly labeled items) and the time needed for creating one label measured in seconds. 
To avoid bias from learning effects \cite{sc20}, we adopted a between-group design. 
Table \ref{table:experimental_groups} provides an overview of the three conditions in the experiment.

\begin{table}[!ht]
\centering
\vspace{2mm}
\caption{Experimental Conditions: Participants were randomly assigned to one of the experimental conditions.}
\label{table:experimental_groups}
\vspace{-2mm}
\small
\scriptsize
\begin{tabularx}{0.9\textwidth}{llX}
\toprule
\textbf{Group} & \textbf{Name} & \textbf{Description}  \\
\midrule
Group 1     & Control Group     &  Participants were asked to label a set of 50 images without time restrictions on their web browsers.  \\
Group 2     &  Rewarded Task Ads     & Participants were asked to play a mobile game for a period of 10 minutes with the objective to score as high as possible. At gameover events, participants were given the option to perform labeling tasks in return for continuing the game with a bonus score of 5 points. \\
Group 3  & Non-Optional Task Ads     & Participants were asked to play the same mobile game from experiment 2 with the same objective. However, after every minute of gameplay, if a gameover event occurs, participants were forced to perform labeling tasks without any incentives. \\ 

\bottomrule
\end{tabularx}
\vspace{0mm}
\newline
\begin{tabularx}{0.9\textwidth}{X} 
  \textit{Notes.} To account for the additional effort of downloading and installing a mobile app participants in Group 2 and Group 3 received and additional USD 0.5 in compensation.
\end{tabularx}
\vspace{2mm}
\end{table}

\subsubsection{Participants}
We recruited 300 participants to evaluate the implemented system.
To avoid selection bias we recruited all participants via Amazon Mechanical Turk. 
Participants were compensated in accordance to the reservation price for MTurk workers \cite{Ma11}.

\subsubsection{Apparatus}
We decided for an image labeling task to evaluate the system because of the frequent need for labeling data in this domain.
The participants of all the three groups were asked to label images from the same dataset.
To be able to calculate the labeling accuracy achieved during the experiment, we obtained all images from the Open Images Dataset \cite{al20}.
The dataset in our experiment consisted of 50 images equally distributed between five classes: \textit{aircraft}, \textit{bird}, \textit{bicycle}, \textit{boat}, and \textit{dog}.
Each class has 5 true positive images and 5 false positive images.
To test the system in a realistic setting we integrated the SDK into a clone of the mobile game Orble\footnote{\url{https://play.google.com/store/apps/details?id=com.chandruscm.orble.android} (last-accessed: 2022-04-21)}.
The objective of the game is to place a gray ball between four orbits without colliding the orange balls. 
The user can switch the orbit of the gray ball by pressing the bottom left or bottom right of the game screen, which moves the gray ball to an adjacent inner orbit or outer orbit. 
The user scores a point by collecting green balls that spawn in random intervals at random orbits. 
The game starts at a score of zero whenever a new game begins. 
A gameover event occurs when the gray ball collides with an orange ball. 
As a result, the game user has to constantly engage with the game by switching the orbits of the gray ball to score and avoid gameover.
The original game is primarily monetized by mobile ads and has been downloaded more than 100,000 times. 
It serves as a representative example of a mobile app that can make use of the proposed solution. 

\subsubsection{Procedure}
\begin{figure*}[!h]
   \vspace{0mm}
   \centering
    \includegraphics[width=0.24\textwidth]{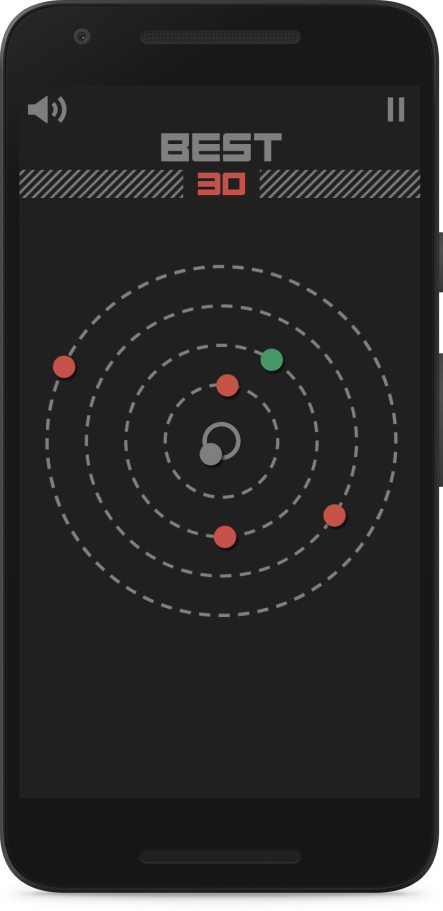}
    \includegraphics[width=0.24\textwidth]{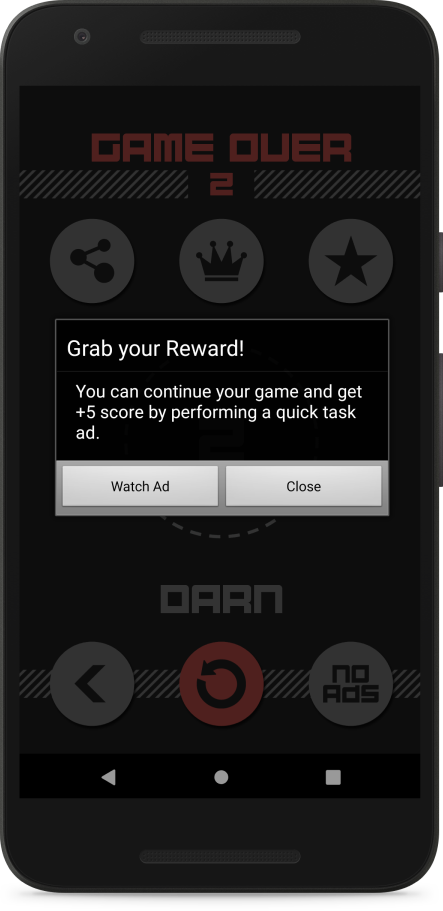}
    \includegraphics[width=0.24\textwidth]{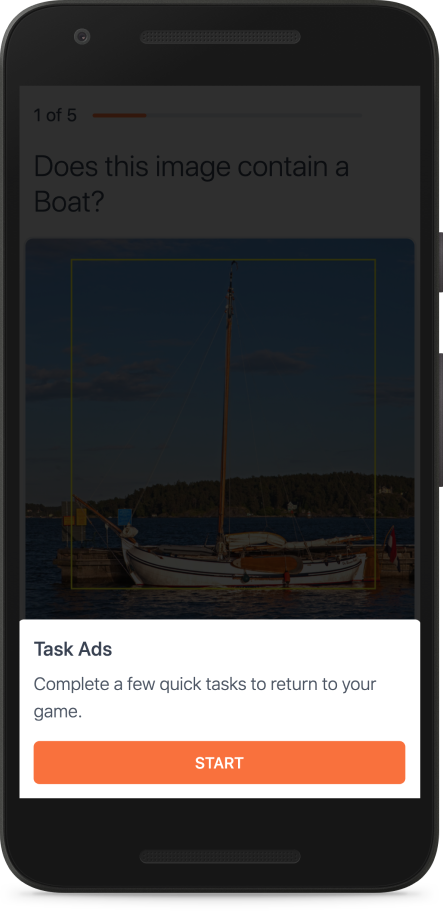}
    \includegraphics[width=0.24\textwidth]{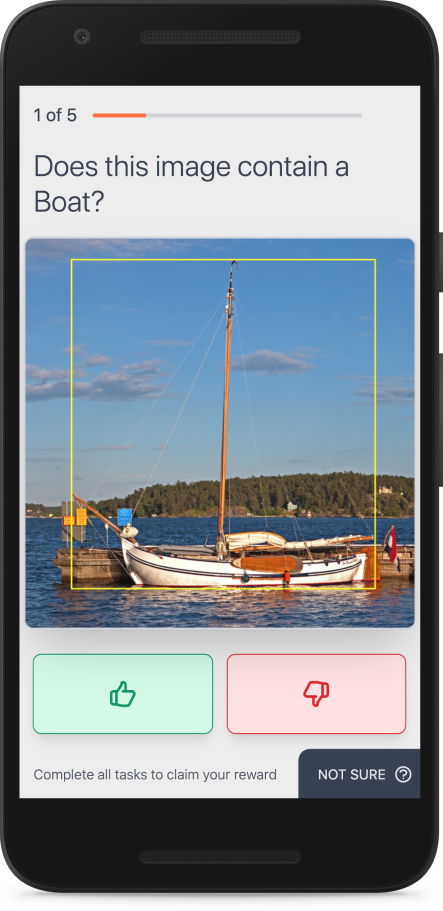}
   \vspace{-0mm}
   \caption{
    Interfaces of the mobile app used for the experiment. From left to right: The game interface, the prompt shown to the \textit{Rewarded Task Ad} group, the prompt shown to the \textit{Non-Optional Task Ad} group, and the labeling interface.
   }
   \label{fig:experiment_interface}
   \vspace{0mm}
\end{figure*}

The procedure differed slightly between control group and task ad groups.
All groups were required to label images and fill out a survey collecting demographic information.
Participants of the control group were asked to label a set of 50 images in their web browser as primary objective without any time restrictions.
Participants of group 2 and 3 were asked to download the mobile game and were given the objective to score as high as possible within 10 minutes.
During the gameplay, the task ads were shown to the participants according to their assigned group (Table \ref{table:experimental_groups}).
The labeling task for all three groups required participants to select whether images were correctly labeled.
The interface presented an image and a corresponding prompt, i.e.\textit{ "Does this image contain a Dog?"}. 
Participants had to choose from three options to label the image, "Yes", "No", and "Not sure". 
The correct option for a true positive image is "Yes" and for a false positive image is "No".
Figure \ref{fig:experiment_interface} provides an overview of the labeling task interfaces in the mobile game environment. 
For the task ad groups, the survey contained additional questions about participants' perceptions.

\subsection{Results}

The experiment was conducted with a total of 300 participants (100 per group) crowdsourced from the MTurk platform.
One participant of experiment 1 and three participants of experiment 3 have been suspected of cheating\footnote{We removed them because they either had the same response for all the questions submitted within a few seconds or they submitted the same survey completion code more than once.}, which is a common occurrence in the platform \cite{ki18}. 
Hence, the corresponding data points were removed from the sample. 
Within the resulting sample (296 participants) 59.6\% were male, and 40.4\% were female.
The average age was 35.7 years. 
The participants were primarily located in the United States of America and India, with 60.9\% and 23.9\% respectively. 

\subsection{Labeling Performance}
Participants of group 1 (control) performed the labeling tasks with the traditional setting, achieving a median success rate of 82\% and a mean success rate of 81.5\%.
Each participant labeled all 50 images in the dataset with a median time of 6.42 seconds per label and a mean time of 7.88 seconds per label.
Tables 7.1 and 7.2 provide an overview of all the collected data.
Participants in group 1 were allowed to use any device of their choice to completing the labeling task.
MTurk workers are less likely to use mobile devices \cite{br14}, hence performing the labeling task with a non-mobile device may indicate a higher task completion time compared to the other groups in which the participants could only use a mobile device.
Group 2 (rewarded task ads) resulted in a similar labeling performance with a median success rate of 80\% and a mean success rate of 76.3\%.
By design of the experiment, participants were given the option to perform a labeling tasks in return for incentives within the game when a gameover event occurs.
72 participants out of 100 chose to perform at least one labeling task, out of which, the median number of labeled images was 17 and the mean number of labeled images was close to 18.
Participants took a median time of 3.99 seconds per label and a mean time of 4.57 seconds per label.
Participants of groups 3 (non-optional task ads) were required to perform labeling tasks within the game at periodic intervals of 1 minute if a gameover event occured. 
We noticed a slight increase in the performance compared to group 2
Participants achieved a median success rate of 84\% and a mean success rate of 80.6\%.
They labeled a median of 25 images and a mean close to 27 images, with a median time of 3.32 seconds per label and a mean time of 4.38 seconds per label.

\begin{table}[!ht]
\vspace{0mm}
\caption{Descriptive statistics per experimental condition.}
\label{table:experiment_results}
\vspace{-2mm}
% \small
\scriptsize
\begin{tabular}{lrrrrrrrrrrrrrrr}
		\toprule
		\multicolumn{1}{c}{} & \multicolumn{3}{c}{\textbf{No. Images Labeled}} & & \multicolumn{3}{c}{\textbf{No. Correct Labels}} & & \multicolumn{3}{c}{\textbf{Success Rate}} & & \multicolumn{3}{c}{\textbf{Time per Label}} \\
        \cmidrule[0.4pt]{1-16}
        % \textbf{Group} & 1 & 2 & 3 & & 1 & 2& 3 & & 1 & 2& 3 & & 1 & 2& 3 \\
        \textbf{Group} & \textbf{c} & \textbf{r} & \textbf{n} & & \textbf{c} & \textbf{r} & \textbf{n} & & \textbf{c} & \textbf{r} &\textbf{ n} & & \textbf{c} & \textbf{r} & \textbf{n} \\
		\cmidrule[0.4pt]{1-16}
		Participants & $99$ & $72$ & $97$ & & $99$ & $72$ & $97$  & & $99$ & $72$ & $97$  & & $99$ & $72$ & $97$  \\
		Missing & $0$ & $28$ & $0$ & & $0$ & $28$ & $0$ & & $0$ & $28$ & $0$ & & $0$ & $28$ & $0$  \\
% 			Median & $50.000$ & $17.000$ & $25.000$  & & $41.000$ & $12.000$ & $22.000$ & & $6.420$ & $3.990$ & $3.320$  & & $0.820$ & $0.800$ & $0.840$  \\
		Median & $50.00$ & $17.00$ & $25.00$  & & $41.00$ & $12.00$ & $22.00$  & & $0.82$ & $0.80$ & $0.84$  & & $6.42$ & $3.99$ & $3.32$  \\
		Mean & $50.00$ & $17.96$ & $26.77$ & & $40.75$ & $13.71$ & $21.42$ & & $0.82$ & $0.76$ & $0.81$ & & $7.88$ & $4.58$ & $4.39$  \\
		Std. Deviation & $0.00$ & $12.54$ & $6.85$ & & $4.92$ & $9.64$ & $6.02$  & & $0.10$ & $0.22$ & $0.14$ & & $5.29$ & $3.39$ & $3.41$  \\
		Minimum & $50.00$ & $1.00$ & $15.00$ & & $2.10$ & $0.00$ & $1.68$ & & $50.00$ & $1.00$ & $15.00$ & & $2.10$ & $0.00$ & $1.68$  \\
		Maximum & $50.00$ & $46.00$ & $50.00$ & & $49.00$ & $37.00$ & $40.00$ & & $0.98$ & $1.00$ & $1.00$ & & $27.16$ & $27.34$ & $22.99$  \\
		\bottomrule
\end{tabular}
\vspace{0mm}
\begin{tabularx}{1.0\textwidth}{X} 
  \textit{Notes.} Group abbreviation as follows: c = control group, r = rewarded task ads group, n = non-optional task ad group. Labeling tasks in the rewarded tasks ads group were voluntary and 28 participants did not choose to start a labeling task and instead just continued playing the game.
\end{tabularx}
\end{table}

\subsubsection{Statistical Tests}
\begin{figure*}[!h]
   \vspace{0mm}
   \centering
    \includegraphics[width=0.49\textwidth]{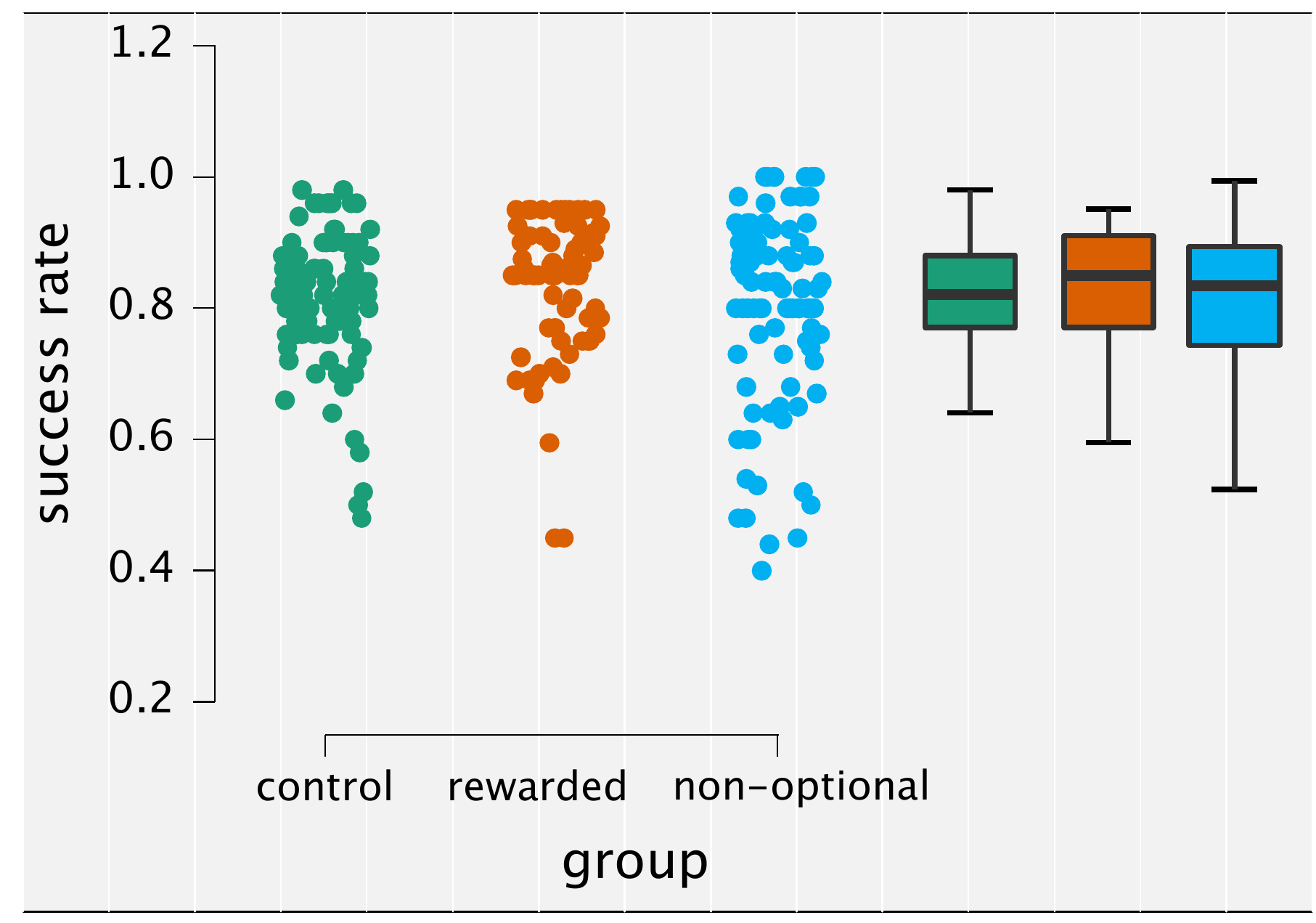}
    \includegraphics[width=0.49\textwidth]{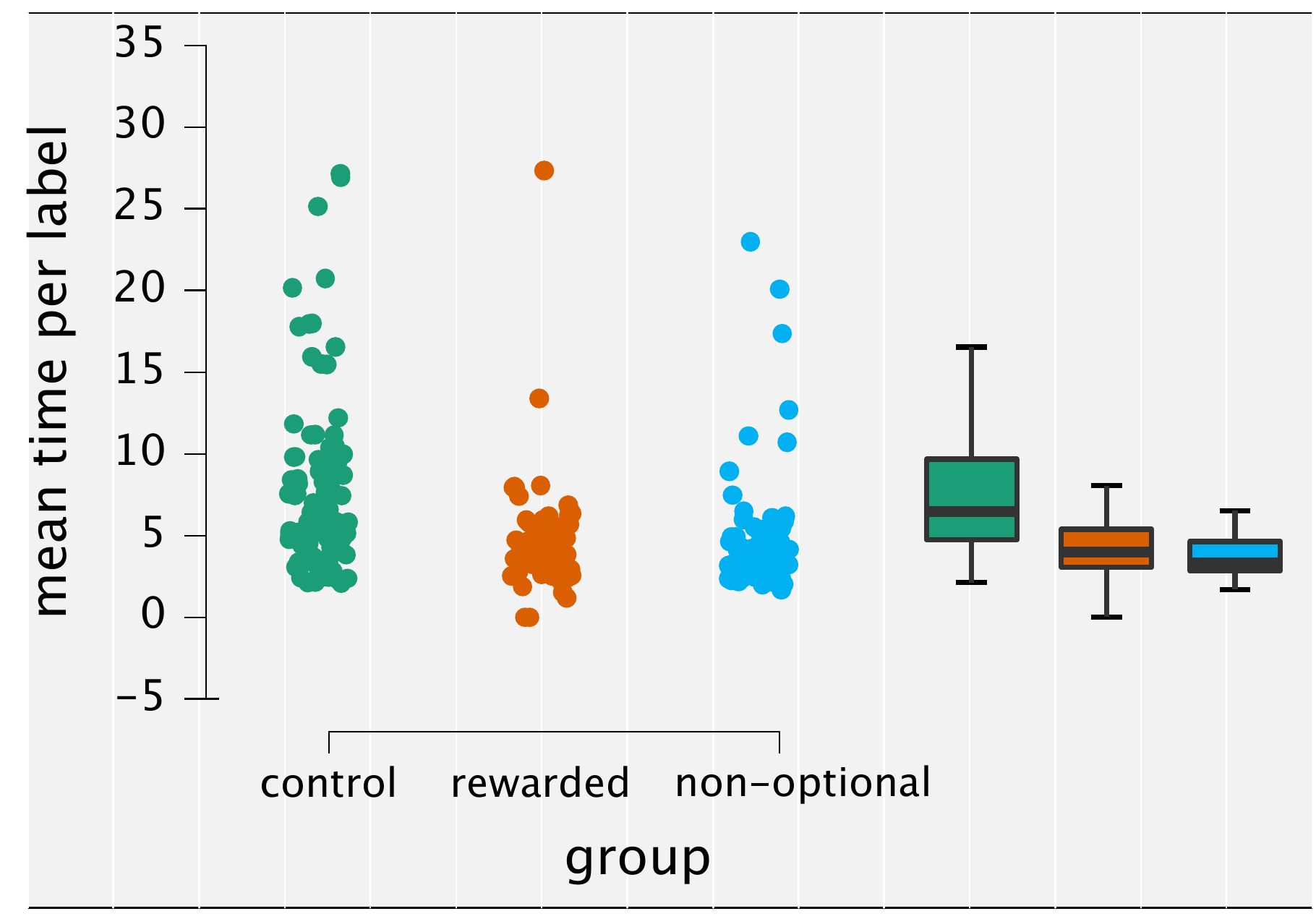}
   \vspace{-0mm}
   \caption{
    The success rate (labeling accuracy) and mean time per label per participant for each of the different experimental conditions.
   }
   \label{fig:label_accuracy}
   \vspace{0mm}
\end{figure*}

To understand the feasibility of our approach for generating high-quality labels for datasets, we looked specifically at the success rate and the time per label produced by each of the groups. 
Figure \ref{fig:label_accuracy} provides an overview of the success rate and the mean time per label for the different experimental conditions.
All statistical tests are conducted with $ \alpha = 0.05$ as threshold for statistical significance.
With a sufficiently large sample size per group (>30), the central limit theorem allows us to assume a normal distribution for all following statistical tests \cite{central2008}.

\paragraph{Labeling Success Rate}
The success rate is calculated as the number of correctly labeled images, divided by the total number of images labeled by each participant.
Descriptive mean success rates differ slightly between the three groups (cf. Table \ref{table:experimental_groups}). 
To understand whether these differences are significant we performed statistical tests.
A Levene's-Test \cite{levene1961robust} showed significant difference in variance between the groups $(F(2,265)=18.021, p<0.001)$, violating the homogeneous variance assumption required to use ANOVA~\cite{faraway2002practical, delacre2019taking}.
Hence, we used Welch's ANOVA, which relaxes the homogeneity of variance assumption~\cite{delacre2019taking}.
Testing with Welch's ANOVA showed that the success rates did not differ significantly between the groups $(F(2,145.352)=1.796, p=0.170)$. 
Since Welch's ANOVA can only state the existence of a difference, we used additional pair-wise post-hoc analysis between the groups to check for robustness. 
We used a Games-Howell post-hoc test as it is suited for comparing groups with unequal variances \cite{games1976pairwise, ruxton2008time}.
The analysis confirms that there is no significant difference between the groups (control-rewarded: $p=0.146$, control-non-optional: $p=0.856$, rewarded-non-optional: $p=0.324$).
In simple words, no difference in the success rates between the groups was found.

\paragraph{Time per Label}
The average time per label is calculated as the total time spent labeling images divided by the total number of images labeled by each participant.
Descriptive mean times per label differ between the three groups, particularly between control group and the task ad groups (cf. Table \ref{table:experimental_groups}). 
To understand whether these differences are significant we performed statistical tests.
A Levene's-Test \cite{levene1961robust} showed significant difference in variance between the groups $(F(2,265)=11.332, p<0.001)$, violating the homogeneous variance assumption required to use ANOVA~\cite{faraway2002practical, delacre2019taking}.
Hence, we used Welch's ANOVA, which relaxes the homogeneity of variance assumption~\cite{delacre2019taking}.
Testing with Welch's ANOVA showed that the mean time per label differed significantly between the groups $(F(2,169.923)=16.563, p<0.001)$. 
Since Welch's ANOVA only states the existence of a difference, we conducted an additional pair-wise post-hoc analysis between the groups. 
We used a Games-Howell post-hoc test as it is suited for comparing groups with unequal variances \cite{games1976pairwise, ruxton2008time}.
The analysis confirms that there is a significant difference between the groups (control-rewarded: $p<0.001$, control-non-optional: $p<0.001$, rewarded-non-optional: $p=0.932$).
In simple words, the groups exposed to task ads needed less time per label when compared to the control group.

\subsection{System Performance}
The prototypical implementation of the TruEyes system was able to handle the live experiments without undergoing any failure. Between the experiments, the app users interacted with TruEyes system 4477 times, and collectively spent nearly 5 hours to label images. TruEyes Client, the subsystem responsible for rendering the labeling task for the app user, is essentially a web app as mentioned in the implementation section. 
On average, the TruEyes Client managed to load the interactive labeling tasks for an app user in 1.8 seconds. 

\subsection{Perception of Task Ads}
Participants of group 2 and 3 were asked additional questions to evaluate how they perceived task ads.
Participants had to answer on a 5-point Likert scale with responses ranging from one to five (1 = Strongly Disagree, 5 = Strongly Agree).
Overall, we observed that participants had a good understanding of the task, they found it easy to complete and tried to answer correctly, which is also evident from the labeling performance.
Participants also preferred task ads over normal ads. Table 7.3 shows the arithmetic mean of the survey responses.\newline

\begin{table}[!ht]  
  \centering
  \scriptsize
  \caption{Perception of Task Ads: Arithmetic mean of the survey responses on 5-point Likert scale.}
    \begin{tabular}{p{0.05\textwidth}p{0.45\textwidth}p{0.20\textwidth}p{0.20\textwidth}}
      \toprule 
      \textbf{No.} & \textbf{Question} & \textbf{rewarded (mean)} & \textbf{non-optional (mean)}\\
      \midrule
      1 & Did you understand the task on the ads? & 4.15 & 4.52 \\
      2 & Did you find the tasks easy to complete? & 4.16 & 4.39 \\
      3 & Did you try to complete the tasks correctly? & 4.42 & 4.47\\
      5 & Do you prefer task ads to normal ads? & 3.80 & 3.80 \\
      \bottomrule
      \addlinespace[1ex]
        \multicolumn{4}{p{0.99\linewidth}}{\textit{Note.} Responses were recorded on 5-point Likert scale (5 = Strongly Agree, 1 = Strongly Disagree). Differences between groups are not statistically significant, except for Question 1 ($t(195)=3.243, p=0.001$).} \\
  \end{tabular}
\end{table}

\subsubsection{Qualitative Perceptions of Task Ads}
We also collected qualitative responses of how task ads were perceived by participants. 
We asked them in an open-ended question: \textit{"If you think about having task ads instead of normal ads, what comes to your mind?}. We received in total 100 responses. 
We used the collected data and clustered the responses to find common themes. 
The summary of the qualitative comments is provided below.

\begin{itemize}
  \item \textbf{Interesting and Engaging:} Twenty participants found the task ads to be interesting, owing to task ads being a new experience and fun to perform. 
  Two participants also expressed their preference for traditional mobile ads due to the fact that they do not need to be inactive until the normal ad stops showing.

  \item \textbf{Less Annoying:} Five participants found the task ads to be less annoying compared to normal ads. Two participants reported the intrusive nature of normal ads that play music.

  \item \textbf{Shorter Duration:} Ten participants preferred the shorter duration of task ads in comparison to normal ads such as video ads which can typically last for 30 seconds. Three participants found the task ads to be a quicker way to get back to their game.

  \item \textbf{Taxing:} Two participants shared that they do not wish to have any type of ads. One participant further expressed that he much prefers a break during mobile ads and not a task.

\end{itemize}

\noindent
Additionally, presented below are a few descriptive responses shared by the participants.
\begin{itemize}

  \item \textit{"I like that the task ads were, for the most part, easy. If I didn't try to rush through them in order to get back to the game quicker, they were easy and took 5 to 10 seconds or so. Watching the video ads while playing games can be extremely intrusive to one's gameplay, and they rarely take less than 30 seconds to 1 minute. Sometimes, constant video ads will be so intrusive and disruptive to gameplay that I will uninstall the game and give up on it completely, even if I like it. I preferred the task ads because they were quicker."}

  \item \textit{"What comes to my mind when I think about task ads is like a quick assignment you would need to do in order to continue using a website, mobile app etc. It would be something easy to do that wouldn't contain a lot of critical thinking."}

  \item \textit{"I much prefer task ads that give me something to do while waiting. I prefer to use my time productively. I usually think of recaptcha or small game like task ads."}

\end{itemize}

\section{Discussion}
From the results of the online experiments, we observe that the success rate and the time per label achieved by the TruEyes data labeling approach is comparable to the baseline set by the traditional labeling approach. Furthermore, the responses from the qualitative survey indicate that the majority of app users have a general reluctance towards current mobile ads. App users are open to new experiences like TruEyes task ads to keep themselves engaged during ad breaks. With these findings, we believe that the TruEyes approach has the potential to mobilize app users for performing crowdsourced labeling tasks. While our evaluation focused on labeling quality and system performance, for the TruEyes system to be commercially viable and competitive to other crowdsourcing platforms, the cost per label and the representativeness of the participants remain to be evaluated in future studies.
In the following we discuss the result of our evaluation and its implications for future research.

\paragraph{Microtasks Use Cases Beyond Image Labeling}
In this paper, we explored data labeling microtasks of images with the TruEyes system, as it has become a popular crowdsourcing task due to the recent rise in machine learning applications \cite{Barbosa2019Rehumanizing}.
Specifically, the labeling performance of the TruEyes system was evaluated only on image datasets. Although the image labeling tasks proved to be a great fit for initial exploration, in reality, depending upon the machine learning application, several kinds of labeling needs may arise, for e.g. text transcription, audio transcription, sentiment analysis, image categorization, video annotation etc. \cite{vo13, ca19, Barbosa2019Rehumanizing}, providing opportunities to explore other kinds of data labeling tasks with the TruEyes system. Apart from data labeling, the TruEyes system is also suitable for crowdsourcing tasks such as, participating in short surveys, market studies and contributing to location based projects, additionally taking advantage of the touch input and sensors present in modern day mobile devices. However, we acknowledge that not every task might be equally suited for microtasks. Certain tasks may require longer durations of undivided attention from the participants, while others may require the participants to recollect previously made available information. In these instances, the ad-hoc app users crowdsourced by TruEyes might not be reliable. The form-factor of a mobile device poses additional constraints on the type of tasks that can be executed. Therefore, further research is necessary to understand what types of crowdsourcing tasks are well suited for the TruEyes system.

\paragraph{Ensuring Quality Control}
Quality control is the one of the major challenges with crowdsourcing \cite{al13}. Platforms such as MTurk employ various quality assurance techniques. On a basic level, requesters have the ability to choose workers based on their reputation as well as their qualification \cite{Ma11}. However, it is unreasonable to expect all workers to perform tasks diligently all the time. Some workers make use of automated bots to complete multiple tasks at the same time, while others find ways to game the system without spending much effort \cite{ch19}. At the same time, some workers may not posses the skills and expertise necessary to complete certain tasks \cite{qu11}. Therefore, it is inevitable that quality control issues will also affect the TruEyes system. Over the years, researchers have come up with different strategies to tackle this problem in crowdsourcing. One such strategy is incorporating explicitly verifiable questions as part of the task \cite{ki08}. For the TruEyes image labeling tasks, this means that every set of labeling tasks would include images for which the label is already known. Another strategy involves taking advantage of expert workers in the pool. Naturally, some users would perform better than others, eventually becoming experts in certain tasks. Frameworks such as ELICE \cite{kh11} make use of a few number of experts to improve the performance of a bigger pool of labeling workers. However, this requires contextual tracking of a mobile app user's performance to identify their areas of expertise. On the other hand, some researchers believe that it impossible to weed out all the adversarial workers, for which they propose error-embracing crowdsourcing models that can tolerate small amounts of error while retaining quality \cite{kr16}.

\paragraph{Bias and Ethics In Crowdsourcing}
One of the major benefits of crowdsourcing platforms like MTurk is the access to a large sample of human workers who are persistently available \cite{Ma11}. It has allowed researchers to overcome the statistical barriers associated with small sample sizes \cite{ki08}. But the mechanics of the MTurk platform implicitly introduces the potential for sampling bias, as workers have the freedom to choose tasks based on monetary compensation, completion times, or other interests and incentives \cite{fo17, al13}. The rising popularity of crowdsourcing platforms has also taken a toll on workers, with issues such as underpayment and difficulties finding appropriate crowdsourcing work \cite{Barbosa2019Rehumanizing}. On the contrary, the TruEyes approach relies on the mobile app ecosystem as its distribution method. By integrating task ads into day-to-day apps and games, we capture responses from app users who might otherwise not participate in crowdsourcing platforms. With the right threshold of apps integrating the TruEyes task ads, we believe that the sampling bias associated with traditional crowdsourcing can be minimized. As app users are intrinsically motivated to engage with their apps, we argue that the non-monetary rewards associated with TruEyes task ads eliminate the notion of crowdsourced workers being treated as a computational resource.

\paragraph{A More Productive Way For App Monetization}
With more and more digital products and services being offered for free on the Internet, ads have become a vital financial component for monetization. Advertisers derive the most value out of ads when they have fine-grained visibility and contextual understanding of the viewers of their ads, which helps target the right ad for the right audience. Over the years, advertisement publishers have adopted aggressive measures to collect user data, raising privacy concerns \cite{pa17}. With the growing awareness on data privacy and introduction of regulations such as GPDR (General Data Protection Regulation), it has become increasingly cumbersome for advertisers to collect user data \cite{le20}. Notably, in light of growing privacy concerns, Apple recently introduced a mandatory opt-in system for enabling tracking for apps on their mobile operating system iOS with version 14 \cite{ko22}. We believe that adopting TruEyes as an alternative monetization strategy can help change users' perception towards free products and services on the Internet. The TruEyes system facilitates an open exchange of value, where users perform crowdsourcing work in the form of microtasks in exchange for accessing free products and services.\\ 

\noindent
As technology continues to grow, the average user will have more number of interactions with digital systems, and we believe the duration of these interactions will get even shorter overtime. Our exploration with TruEyes task ads demonstrated how capturing the user's attention for a short period of 30 seconds in between app usage could provide value if the incentives are aligned. We believe that leveraging these short interactions for performing crowdsourcing or other forms of colloborative work opens new opportunities with interesting areas to further expand HCI research.

\section{Conclusion}
We explored a novel approach for crowdsourcing data labeling tasks by making use of app users. Particularly, we wanted to verify whether app users can serve as a viable pool of crowdsourced workers and overcome the shortcomings of traditional crowdsourcing. To realize this, we developed TruEyes, which crowdsources app users to participate in data labeling tasks while they are using an app, referred to as task ads. To evaluate mobile app users, we implemented a prototype for Android mobile users and conducted a randomized online experiment with N=296 participants, out of which, 99 participants performed the labeling task with a traditional crowdsourced setting to establish a performance benchmark. The results indicate that app users performed similar to workers that undertook the traditional labeling approach by achieving a median success rate of 80\% and 84\% in two independent experiments, while the benchmark was 82\%. Workers in traditional crowdsourced platforms lose motivation due to the repetitive nature of labeling tasks. Our post-experiment survey indicates that the majority of participants felt task ads were more engaging and generally preferred task ads over normal ads. 
We conclude with a discussion reflecting on the potential of microtasks as alternative form of app monetization.

\bibliographystyle{unsrt}  
% \bibliography{references}  %%% Remove comment to use the external .bib file (using bibtex).


\end{document}